\documentclass[11pt]{article}
\usepackage{graphicx}
\usepackage{amsmath,amsxtra,amssymb,latexsym, amscd}
\usepackage[mathscr]{eucal}

\makeindex

\numberwithin{equation}{section}
\begin{document}
\parindent=1.05cm %Khong thay doi
\setlength{\baselineskip}{12truept} \setcounter{page}{1}
\makeatletter
\renewcommand{\@evenhead}{\@oddhead}
\renewcommand{\@oddfoot}{}% empty footer
\renewcommand{\@evenfoot}{\@oddfoot}
\renewcommand{\thesection}{\arabic{section}.}
\renewcommand{\thesubsection}{\thesection\arabic{subsection}.}
\renewcommand{\theequation}{\thesection\arabic{equation}}
\@addtoreset{equation}{section}
\begin{center}
\vspace{10cm}
{\bf FUNCTIONAL INTEGRAL METHOD FOR POTENTIAL SCATTERING AMPLITUDE IN QUANTUM MECHANICS}\\
\vspace{0.5cm}
\small{
Cao Thi Vi Ba$^{1,a}$,Do Thu Ha$^{1,2,b}$, Nguyen Nhu Xuan$^{3,c}$\\
\vspace{0.5cm}
{$^1$\it Department of Theoretical Physics, Hanoi University of Science, Hanoi, Vietnam.} \\
{$^2$\it University of Natural Resources and Environment, Hanoi, Vietnam.} \\
{$^3$\it Department of Physics, Le Qui Don Technical University, Hanoi, Vietnam.}
\footnote{email:{{$^a$\it caoviba@yahoo.com}, {$^b$\it thuhahunre@gmail.com},{$^c$\it xuannn@lqdtu.du.vn}.}}}
\end{center}

\vspace{0.5cm}
\baselineskip=18pt
\bigskip
\textbf{Abstract}: The functional integral method can be used in quantum mechanics to find the scattering amplitude for particles in the external field. We will obtain the potential scattering amplitude form the complete Green function in the corresponding external field through solving the Schrodinger equation, after being separated from the poles on the mass shell, which takes the form of an eikonal (Glauber) representation in the high energy region and the small scattering angles. Consider specific external potentials such as the Yukawa or Gaussian potential, we will find the corresponding differential scattering cross-sections.\\

\textbf{Keywords:} Eikonal scattering theory, effective theory of quantum gravity, quasi-potential equation and modified perturbation theory.
\vspace{0.5cm}

\section{Introduction}
\indent The eikonal approximation for the potential scattering amplitude can be applied when solving the Schrodinger equation or when expanding the perturbation theory series of the scattering amplitude in the Born approximation [1]. These two approaches give us the basis for applying the eikonal approximation in quantum field theory, at a region where the concept of potential cannot be used.\\
\indent In this paper we would like to introduce a new method, the functional integral method to find the Green function of a particle from the Schrodinger equation in the external field. The eikonal approximation here is equivalent to the straight line approximation, which is used to compute the functional integrals as it occurs. The advantage of this new approach is that it can be extended to compute the leading term and the first-order correction term in the asymptotic scattering amplitude at Planck energies and the first-order correction in quantum linear gravity theory [2-6] and effective quantum gravity theory [7].\\
\indent The content of the paper is presented as follows: In section 2, we will briefly introduce how to represent the Green function of the particle in the external field in the form of functional integral from Schrodinger's equation and how to separate the poles from the Green function of the particle in the mass shell, to find the potential scattering amplitude. The method of calculating the functional integral by using the straight line approximation and consideration of the asymptotic shape of the potential scattering amplitude at high energy region and small scattering angle is presented in section 3. The conditions of the potential, energy of the particle and the scattering angles for which this approximation can be used are discussed in this section. In section 4, we consider the differential scattering cross-section given specific external potentials such as Yukawa and Gauss potentials. The concluding section is devoted to the resulting systems and discusses the possibility of extending this approach to more complex problems in subsequent studies. Here the atomic unit system $\hbar  = c = 1$ and metric Feynman are used.
\section{Two-particle quasi-potential equation in an operator form}
\indent The amplitude of the particle scattering in the external field can be found through solving the Schrodinger equation. First, we solve the integral equation corresponding to Schrodinger equation to find the Green function of the particle in the external field [1,8].
\begin{equation}\label{2.1}
   \left(E + \frac{\hbar ^2}{2m}{\vec\nabla}^2 - V(\vec{r})+ i\varepsilon\right)G\left(\vec{r},\vec {r'}\right) = \delta^3\left(\vec {r}- \vec {r'}\right).
\end{equation}
Here $E$ in \eqref{2.1}was replaced by $E + i\varepsilon$ to get the Green function, which only contains divergence expressions when $r \to \infty $.\\
Applying the Feynman, Fock representation to the inverse operator representation in an exponential form, we can write the solution of equation \eqref{2.1} in operator form as follow
\begin{equation}\label{2.2}
\begin{split}
G\left( {\vec r,\vec r'} \right) =& \left(E +\frac{\hbar^2}{2m}{\vec\nabla}^2 - V(\vec{r}) + i\varepsilon\right)^{-1}\delta^3(\vec r - \vec {r'})\\
 =&- i\int_0^\infty d\tau \rm{exp}\left\{ i\int_0^\tau d\xi\left(E - \frac{{\vec p}^2(\xi)}{2m} - V(\vec r,\xi) + i\varepsilon \right)\right\}\delta ^3\left(\vec r - \vec {r'}\right),
\end{split}
\end{equation}
where $\vec p\left( \xi  \right) =  - i\hbar {\vec \nabla _x}\left( \xi  \right)$ is the momentum operator.\\
The exponential term in the \eqref{2.2}, which contains non-commuting operators $\vec \nabla _\xi ^2$  and $V\left( {\vec r,\xi } \right)$ is considered as $ T_{\xi}$ - exponent, where the ordering subscript has the meaning of the proper time divided by mass $ m $. All operators in \eqref{2.2} are assumed to be commuting functions that depend on the parameters $\xi$. In the power of exponential there is a second derivative differential derivative $ {{\vec p}^2}\left( \xi  \right) =  - {\hbar ^2}\vec \nabla _x^2\left( \xi  \right)$. However, the transition from $ T_\xi $ - exponent to ordinary operator expression (“disentangling” the differentiation operators in the argument  of the exponent function by terminogy of Feynman) cannot be performed without the series  expansion with respect to an external field  $ V\left( {\vec r,\xi } \right) $. But one can lower the the power of the operator ${{\vec p}^2}\left( \xi  \right)$ in \eqref{2.2} by using the following formal transformation that contains an integral function of three dimensions [6].
\begin{equation*}
\begin{split}
\exp \left\{{- \frac{i}{{2m}}\int_0^\tau {{{\vec p}^2}(\xi )d\xi } } \right\} =& \exp \left[ { - i{{\int_0^\tau {\left( {\frac{{\vec p(\xi )}}{{\sqrt {2m} }} + \vec \nu (\xi )} \right)} }^2}d\xi } \right]\exp \left[ {i\int_0^\tau {\left( {2\vec \nu (\xi )\frac{{\vec p(\xi )}}{{\sqrt {2m} }} + {{\vec \nu }^2}(\xi )} \right)d\xi } } \right]\\
=&{C_\nu }\int {\prod_\eta {{d^3}\nu (\eta)} } \exp \left\{ {i\int_0^\tau {{{\vec \nu }^2}(\eta)d\eta  + 2\frac{i}{{\sqrt {2m} }}\int_0^\tau  {\vec p(\xi)\vec \nu (\xi)d\xi } } } \right\},
 \end{split}
\end{equation*}
\begin{equation}\label{2.3}
\begin{split}
G(\vec r,\vec r') =&- i\int_0^1 {d\tau \exp \left\{ {d\xi \left( {E + i\varepsilon } \right)} \right\}} {C_\nu }\int {\prod\eta {{d^3}\nu (\eta)}} \exp \left\{ {i\int_0^\tau  {{{\vec \nu }^2}(\eta)d\eta}}\right\}\times \\
 =&- i\int_0^\infty  {d\tau \exp \left\{ {i\tau(E + i\varepsilon)}\right\}} {C_\nu }\int {\prod_\eta  {{d^3}\nu (\eta)}} \exp \left\{ {i\int_0^\tau  {{{\vec \nu }^2}(\eta)d\eta } } \right\} \times \\
 &\times \exp \left( {i\sqrt {\frac{2}{m}} \int_0^\tau  {\vec p(\xi )\vec \nu (\xi )d\xi } } \right)\exp \left[ {- i\int_0^\tau {V(\vec r,\xi )d\xi } } \right]{\delta ^{(3)}}(\vec r - \vec r').
\end{split}
\end{equation}
Where $\exp \left( {i\sqrt {\frac{2}{m}} \int_0^\tau {\vec \nu (\xi )\vec p(\xi )d\xi } } \right)$ translation operator when moving coordinates by  $\sqrt {\frac{2}{m}} \int_0^\tau  {\vec \nu (\xi )d\xi } $, "Rearrange" the operator expression. Then the Green function for Schrodinger equation in the external potential field, can be written as
\begin{equation}\label{2.4}
\begin{split}
&G(\vec r,\vec r')=  - i\int_0^\infty {d\tau \exp \left\{ {i\tau(E + i\varepsilon)} \right\}} {C_\nu }\int {\prod_\eta  {{d^3}\vec \nu(\eta)} } \\
&\exp \left\{ {i\int\limits_0^\tau  {{{\vec \nu }^2}\left( \eta  \right)d\eta }  - i\int_0^\tau  V\left( {\vec r + \hbar \sqrt {\frac{2}{m}} \int_\xi ^\tau \vec v (\eta)d\eta} \right)d\xi } \right\} \times{\delta ^{(3)}}\left({\vec r + \hbar \sqrt {\frac{2}{m}} \int_0^\tau  \vec v(\eta)d\eta - \vec r'}\right).
 \end{split}
\end{equation}
\eqref{2.4} can be changed into Feynman integral (path integral) by changing the variables $\tau $  into $\frac{t}{\hbar }$  and  $\vec v\left( {\frac{\eta }{\hbar }} \right)$ into  $\vec \omega (\eta)$, we obtain
\begin{equation*}
\begin{split}
G(\vec r,\vec r')=&  - i\int\limits_0^\infty  {\left( {\frac{t}{\hbar }} \right)} {e^{i(E + i\varepsilon )\frac{t}{\hbar }}}{C_\omega }\int {\prod\limits_\eta  {{d^3}\vec \omega (\eta )\exp \left\{ {i\int_0^{{t\mathord{\left/{\vphantom {t \hbar }} \right.
 \kern-\nulldelimiterspace} \hbar }} {{\omega ^2}(\hbar \eta )d(\hbar \eta )\frac{1}{\hbar }} } \right\}} }  \times \\
 &\times \exp \left[ { - i\int\limits_0^{{t \mathord{\left/{\vphantom {t \hbar }} \right.
 \kern-\nulldelimiterspace} \hbar }} {V\left( {\vec r + \hbar \sqrt {\frac{2}{m}} \int_0^{{t \mathord{\left/
 {\vphantom {t \hbar }} \right.
 \kern-\nulldelimiterspace} \hbar }} {\vec \omega (\hbar \eta )d(\hbar \eta )\frac{1}{\hbar }} } \right)d(\xi \hbar )} } \right]\\
 &\times{\delta^3}\left( {\vec r + \hbar \sqrt {\frac{2}{m}} \int_0^{{t \mathord{\left/{\vphantom {t \hbar }} \right.
 \kern-\nulldelimiterspace} \hbar }} {\vec \omega (\hbar \eta )d(\hbar \eta )\frac{1}{\hbar } - \vec r'} } \right).
\end{split}
\end{equation*}
Set $ \hbar \xi  = t',\hbar \eta  = t''$, we have
\begin{equation}\label{2.5}
\begin{split}
G\left( {\vec r,\vec r'} \right) = & - \frac{i}{\hbar }\int\limits_0^\infty  {\left( {\frac{t}{\hbar }} \right)} {e^{\frac{i}{\hbar }(E + i\varepsilon )t}}{C_\omega }\int {\prod\limits_\eta  {{d^3}\vec \omega (\eta ) \times } } \\
&\times \exp \left\{ {\frac{i}{\hbar }\int\limits_0^t {{\omega ^2}(t')d(t') - \frac{i}{\hbar }\int\limits_0^t {V\left( {\vec r + \sqrt {\frac{2}{m}} \int\limits_{t'}^t {\vec \omega (t'')d(t'')} } \right)d(t')} } } \right\}\\
&\times{\delta ^3}\left( {\vec r + \sqrt {\frac{2}{m}} \int\limits_0^t {\vec \omega (t')d(t') - \vec r'} } \right).
\end{split}
\end{equation}
Now we will take the integral of the trajectory $ x\left( {\vec t} \right) $, where  $ x\left( {\vec t} \right) $ is determined by equation
\begin{equation}\label{2.6}
\sqrt {\frac{2}{m}} \int_{t'}^t \vec \omega \left( {t''} \right)dt'' =  - \vec r + \vec x\left( {t'} \right).
\end{equation}
The Jacobian of this transformation does not depend on derivative of the new functional variable $ x\left( {\vec t} \right) $
\[{D^{ - 1}} = det\left\| {\frac{{\delta (\vec x\left( {t'} \right)}}{{\delta \left( {\vec \omega \left( {t} \right)} \right)}}} \right\| = det\left\| {\sqrt {\frac{2}{m}} \mathop \int_{t'}^t \delta \left( {t - t'''} \right)dt'''} \right\| = det\sqrt {\frac{2}{m}} \left\| {\theta \left( {t - t''} \right)\theta \left( {t'' - t} \right)} \right\|\]
Thus, it is a certain constant that we can include it in the normalized constant $ c_{x} $. Let $F(t^{''})$ be a primitive of $\omega (t'')$ then
\begin{equation*}
\begin{split}
 x(t') =& \vec r + \sqrt {\frac{2}{m}} \left( {F(t) - F(t')} \right),\\
 \dot \vec x(t') =&  - \sqrt {\frac{2}{m}} \dot F(t') =  - \sqrt {\frac{2}{m}} \omega (t'),\\
 {\omega ^2}(t') =& \frac{{m\dot \vec x(t')}}{2}.
\end{split}
\end{equation*}
Substitue these equations into Eq.\eqref{2.5}, the Green funtion $G\left( {\vec r,\vec r'}\right)$ has the form
\begin{equation}\label{2.7}
G\left( {\vec r,\vec r'} \right) =  - \frac{i}{\hbar }\mathop \smallint \limits_0^\infty  dt{e^{i\frac{t}{\hbar }\left( {E + i\varepsilon } \right)}}{C_x}\int {\prod\limits_{t'} {d\vec x\left( {t'} \right)\exp \left\{ {\frac{i}{\hbar }\mathop \int_0^t dt'\left[ {\frac{{m{{\dot \vec x}^2}\left( {t'} \right)}}{2} - V\left( {\vec x\left( {t'} \right)} \right)} \right]} \right\}} },
\end{equation}
 with the condition
\begin{equation}\label{2.8}
\begin{split}
x(t)=& \vec r +\sqrt {\frac{2}{m}} \int_t^t{\omega (t'')dt'' = \vec r},\\
\vec x(0) =& \vec {r'}.
\end{split}
\end{equation}
The functional integration in \eqref{2.7} is the Feynman integral by path  $\vec x(t)$ of the particle in the exponential term and the expression of this term is the classical action in the external potential field $V\left(\vec x(t)\right)$.

\section{Scattering amplitude in the Eikonal approximation}
The scattering amplitude of the particle scattering in the external field is calculated by the following formula which is related to Green function
 \begin{equation}\label{3.1}
f\left( {\vec k,\vec k'} \right) =  - \frac{{4{\pi ^2}m}}{{{\hbar ^2}}}\left( {E - \frac{{{\hbar ^2}\vec k{'^2}}}{{2m}} + i\varepsilon } \right) < \vec k'{\rm{|}}G - {G_0}{\rm{|}}\vec k > \left( {E - \frac{{{\hbar ^2}{{\vec k}^2}}}{{2m}} + i\varepsilon } \right),
\end{equation}
where $E = \frac{{{\hbar ^2}\vec k{'^2}}}{{2m}} = \frac{{{\hbar ^2}{{\vec k}^2}}}{{2m}}$.\\
In the above formula, the Green function is used in eq.\eqref{2.4} because the delta function here takes into account boundary conditions \eqref{2.8} on particle orbit. To get the scattering amplitude, we need to separate from the difference $(G - {G_0})$ two poles ${\left( {E - \frac{{{\hbar ^2}{{\vec k}^{'2}}}}{{2m}} + i\varepsilon } \right)^{ - 1}}$ and ${\left( {E - \frac{{{\hbar ^2}{{\vec k}^2}}}{{2m}} + i\varepsilon } \right)^{ - 1}}$ so that they can eliminate the terms  $\left( {E - \frac{{{\hbar ^2}{{\vec k}^{'2}}}}{{2m}} + i\varepsilon } \right)\left( {E - \frac{{{\hbar ^2}{{\vec k}^2}}}{{2m}} + i\varepsilon } \right)$ in \eqref{3.1}. To do these, we can perform the following steps:
 i/ convert to momentum representation, ii/ perform the functional transformation
\begin{equation}\label{3.2}
\begin{split}
 {\vec \omega ^2}\left( \eta  \right)= &{\vec v^2}\left( \eta  \right) + \frac{{{\hbar ^2}}}{{2m}}{\vec k^2} + 2\frac{\hbar }{{\sqrt {2m} }}\vec v\left( \eta  \right)\vec k;\\
  \vec \omega \left( \eta  \right) =& \vec v\left( \eta  \right) + \frac{\hbar }{{\sqrt {2m} }}\vec k.
  \end{split}
\end{equation}
The final result, we can find the Green function of the particle in the external field in the momentum representation.

\begin{equation}\label{3.3}
\begin{split}
\left\langle\vec{k'}\rm{|}G\rm{|}\vec k\right\rangle =&G(\vec k,\vec{k'}) =  - i\int_0^\infty d\tau e^{i\left(E - \frac{\hbar^2{\vec k}^2}{2m}+ i\varepsilon\right)\tau} \int{\frac{{d\vec r}}{{{{(2\pi)}^3}}}{e^{i\left(\vec k - \vec{k'}\right)\vec r}}} {C_\omega }\int {\prod\limits_\eta  {d^3\vec \omega(\eta)}} \times\\
 &\times \exp \left\{{i\int_0^\tau {{\vec \omega }^2}\left( \eta  \right)d\eta - i\int_0^\tau  V\left( {\vec r + \frac{{2\hbar }}{{\sqrt {2m} }} \int_{\xi^t}\vec \omega(\eta)d\eta  - \frac{\hbar^2}{2m}\vec k(\tau - \xi)}\right)d\xi} \right\},
\end{split}
\end{equation}
if in eq.\eqref{3.3} with $ V = 0$; ${C_\omega }\int {\prod\limits_\eta  {{d^3}\vec \omega \left( \eta  \right)} } \exp \left\{ {i\mathop \smallint \limits_0^\tau  {{\vec \omega }^2}\left( \eta  \right)d\eta } \right\} = 1$ then
\begin{equation}\label{3.4}
\left\langle {\overrightarrow {k'} {\rm{|}}{G_0}{\rm{|}}\vec k} \right\rangle  = \frac{{{\delta ^3}(\vec k - \vec k')}}{{E - \frac{{{\hbar ^2}{{\vec k}^2}}}{{2m}} + i\varepsilon }}.
\end{equation}
Note that, the above formula coincides with the Green function of the Schrodinger equation for free particles. So we can remove from the total Green function of the particles in external field $ \left\langle {\overrightarrow k {\rm{'|}}G{\rm{|}}\vec k} \right\rangle  $ contribution of the Green function of the free particles $ \left\langle {\overrightarrow k {\rm{'|}}{G_0}{\rm{|}}\vec k} \right\rangle  $  does not contribute to the scattering amplitude, and uses the formula ${e^a} - 1 = a\int\limits_0^1 {{e^{\lambda a}}d\lambda } $,   we obtain: $\exp \left( { - i\int V } \right) - 1 =  - i\int V \int\limits_0^1 {\exp \left( { - i\lambda \int V } \right)d\lambda } $, the result for $\left\langle {\overrightarrow {k'} {\rm{|}}G - {G_0}{\rm{|}}\vec k} \right\rangle$  we obtain
\begin{equation}\label{3.5}
\begin{split}
\left\langle {\overrightarrow k {\rm{'|}}G - {G_0}{\rm{|}}\vec k} \right\rangle  = & - i\mathop \smallint \limits_0^\infty  d\tau \;{e^{i\left( {E - \frac{{{\hbar ^2}{{\vec k}^2}}}{{2m}} + i\varepsilon } \right)\tau }}\int {\frac{{d\vec r}}{{{{\left( {2\pi } \right)}^3}}}{e^{i\left( {\vec k - \overrightarrow {k'} } \right)\vec r}}{C_\omega }\int {\prod\limits_\eta  {{d^3}\vec \omega \left( \eta  \right){e^{^{i\mathop \smallint \limits_0^\tau  {{\vec \omega }^2}\left( \eta  \right)d\eta }}}}  \times } } \\
& \times \left( { - i} \right)\int\limits_0^1 {V\left( {\vec r + \frac{{2\hbar }}{{\sqrt {2m} }}\mathop \smallint \limits_\alpha ^\tau  \vec \omega \left( \eta  \right)d\eta  - \frac{{{\hbar ^2}\vec k}}{m}\left( {\tau  - \alpha } \right)} \right)d\alpha }  \times \\
&\times \int\limits_0^1 {\exp \left\{ { - i\lambda \mathop \smallint \limits_0^\tau  V\left( {\vec r + \hbar \sqrt {\frac{2}{m}} \mathop \smallint \limits_\xi ^\tau  \vec \omega \left( \eta  \right)d\eta  - \frac{{{\hbar ^2}\vec k}}{m}\left( {\tau  - \xi } \right)} \right)d\xi } \right\}d} \lambda.
 \end{split}
\end{equation}
continue to change variables
\begin{equation*}
\vec r + \hbar \sqrt {\frac{2}{m}} \int_\alpha^\tau \vec \omega (\eta)d\eta - \frac{\hbar^2\vec k}{m}(\tau - \alpha) = \vec x;
\vec \omega (\eta) - \hbar \sqrt{\frac{1}{2m}} \theta(\eta - \alpha)(\vec k - \vec{k'}) = \vec {\omega_1}(\eta).
\end{equation*}
Transform the argument of eq.\eqref{3.5} separately,
\begin{equation*}
  \begin{split}
  &\vec x - \hbar \sqrt {\frac{2}{m}} \int_\alpha^\tau  \vec \omega (\eta)d\eta  + \frac{{{\hbar ^2}\vec k}}{m}(\tau -\alpha) + \frac{2\hbar}{\sqrt{2m}}\int_\xi ^\tau  \vec \omega (\eta)d\eta  - \frac{{{\hbar ^2}\vec k}}{m}(\tau  - \xi)\\
 =&\vec x - \hbar \sqrt {\frac{2}{m}} \int_\alpha ^\xi  \vec \omega (\eta)d\eta  + \frac{\hbar^2}{m}\vec k(\xi  - \alpha)\\
 =&\vec x - \hbar \sqrt {\frac{2}{m}} \int_\alpha ^\xi  \vec \omega (\eta)d\eta  + \frac{\hbar^2}{m}(\vec k\theta(\alpha - \xi) + \vec {k'}\theta  (\xi - \alpha))(\xi - \alpha).
 \end{split}
\end{equation*}
Since the argument in the expression of potential $V$ is integral according to $\xi $ with $\xi$ runs from $0 \to \tau $,\\
when $\xi < \alpha$  then the term $\vec k\theta (\alpha - \xi) + \vec {k'}\theta (\xi  - \alpha) = \vec k$ (before scattering).\\
when $\xi >\alpha$ then the term $\vec k\theta(\alpha - \xi) + \vec {k'}\theta (\xi - \alpha) = \vec {k'}$ (after scattering). we have
\begin{equation*}
  \begin{split}
 &\vec r = \vec x - \hbar \sqrt {\frac{2}{m}}\int_\alpha ^\tau \vec \omega (\eta)d\eta  + \frac{\hbar ^2\vec k}{m}(\tau - \alpha),\\
&{\vec \omega}^2(\eta)+ \frac{\hbar^2}{2m}\theta ^2(\eta - \alpha)(\vec k -\vec{k'})^2 + 2\frac{\hbar}{\sqrt {2m}}\vec \omega (\eta)\theta (\eta - \alpha)(\vec k - \vec{k'}) = \vec \omega _1^2(\eta).
   \end{split}
\end{equation*}
After a series of complex transformations we obtained
\begin{equation}\label{3.6}
\begin{split}
\left\langle {\vec k {\rm{'|}}G - {G_0}{\rm{|}}\vec k} \right\rangle  =  - \int_0^\infty  d\tau \;\exp \left[ {i\left( {E - \frac{{{\hbar ^2}{{\vec k}^2}}}{{2m}} + i\varepsilon } \right)\tau } \right]\exp \left[ {\frac{{i{\hbar ^2}}}{{2m}}\left( {\tau  - \alpha } \right)\left( {{{\vec k}^2} - \vec k{'^2}} \right)} \right] \times \\
 \times {C_{{\omega _1}}}\int {\prod\limits_\eta  {{d^3}{{\vec \omega }_1}\left( \eta  \right)\exp \left[ {i\int\limits_0^\tau  {\vec \omega _1^2(\eta )d\eta } } \right]} \int {\frac{{d\vec x}}{{{{(2\pi )}^3}}}} } \exp \left[ {i(\vec k - \vec k')\vec x} \right]V(\vec x) \times \\
 \times \mathop \smallint \limits_0^1 d\lambda \exp \left\{ { - i\lambda \mathop \smallint \limits_0^\tau  V\left( {\vec x - \hbar \sqrt {\frac{2}{m}} \mathop \smallint \limits_\alpha ^\xi  \vec \omega \left( \eta  \right)d\eta  + \frac{{{\hbar ^2}}}{m}\left( {\xi  - \alpha } \right)\left[ {\vec k\theta \left( {\alpha  - \xi } \right) + \vec k'\theta \left( {\xi  - \alpha } \right)} \right]} \right)d\xi } \right\}.
 \end{split}
\end{equation}
Changing the order of integrating by variables $ \tau $  and $ \alpha $  and assuming that  $\tau  = {\tau _1} + \alpha $
\begin{equation*}
\int_0^\infty {d\tau \int_0^\tau {d\alpha}} = \int_0^\infty d\alpha \int_\alpha^\infty d\tau = \int_0^\infty d\alpha \int_0^\infty d{\tau _1}
\end{equation*}
\begin{equation}\label{3.7}
\begin{split}
&\left\langle {\vec k {\rm{'|}}G - {G_0}{\rm{|}}\vec k} \right\rangle =  - \int_0^\infty  d\alpha \int_0^\infty  {d{\tau _1}} \exp \left[ {i\left( {E - \frac{{{\hbar ^2}{{\vec k}^2}}}{{2m}}} \right)\alpha } \right]\exp \left[ {i\left( {E - \frac{{{\hbar ^2}\vec k{'^2}}}{{2m}}} \right){\tau _1}} \right] \times \\
&\times \exp \left[ {\frac{{i{\hbar ^2}}}{{2m}}\left( {{{\vec k}^2} - \vec k{'^2}} \right){\tau _1}} \right]{C_{{\omega _1}}}\int {\prod\limits_\eta  {{d^3}{{\vec \omega }_2}\left( \eta  \right)\exp \left[ {i\int_{- \alpha }^{{\tau _1}} {\vec \omega _2^2(\eta )d\eta } } \right]}} \\
&\int {\frac{{d\vec x}}{{{{(2\pi )}^3}}}} V(\vec x)\exp \left[ {i(\vec k - \vec k')\vec x} \right]\int_0^1 d\lambda  \times \\
&\times \exp \left\{ { - i\lambda \int_{-\alpha}^{{\tau _1}} V\left( {\vec x - \hbar \sqrt {\frac{2}{m}} \int_0^{{\xi _1}} {{\vec \omega }_2}\left( \eta  \right)d\eta  + \frac{{{\hbar ^2}}}{m}{\xi _1}\left[ {\vec k\theta \left( { - {\xi _1}} \right) + \vec k'\theta \left( {{\xi _1}} \right)} \right]} \right)d{\xi _1}} \right\}.
 \end{split}
\end{equation}
Using the equation $\lim_{x \to {x_0}} ( x - {x_0})\int_0^\infty  {e^{i(x - {x_0})\tau }}{\rm{\Phi }}(\tau)d\tau  = i{\rm{\Phi }}(\infty)$\\
Then the scattering amplitude take the form
\begin{equation}\label{3.8}
\begin{split}
f\left( {\vec k,\vec k'} \right) =  - \frac{{4{\pi ^2}m}}{{{\hbar ^2}}}\mathop \smallint \nolimits^ \frac{{d\vec x}}{{{{\left( {2\pi } \right)}^3}}}V\left( {\vec x} \right){e^{i\left( {\vec k - \overrightarrow {k'} } \right)\vec x}}{C_\omega }\int {\prod\limits_\eta  {d\vec \omega \left( \eta  \right){e^{i\mathop \smallint \limits_{ - \infty }^\infty  {{\vec \omega }^2}\left( \eta  \right)d\eta }} \times } } \\
 \times \mathop \smallint \limits_0^1 d\lambda \exp \left\{ { - i\lambda \mathop \smallint \limits_{ - \infty }^\infty  V\left( {\vec x - \hbar \sqrt {\frac{2}{m}} \mathop \smallint \limits_0^\xi  \vec \omega \left( \eta  \right)d\eta  + \frac{{{\hbar ^2}}}{m}\xi \left[ {\vec k\theta \left( { - \xi } \right) + \overrightarrow {k'} \theta \left( \xi  \right)} \right]} \right)d\xi } \right\}.
\end{split}
\end{equation}
The functional integral in eq.\eqref{3.8} is performed according to the orbits\\ $\vec x\left( t \right) =  - \hbar \sqrt {\frac{2}{m}} \int_0^\xi  {\vec \omega (\eta )d\eta  + \hbar \xi ( {\vec v\theta ( - \xi ) + \vec v'\theta (\xi )} )}$, which satisfied the equation
\begin{equation*}
\frac{{d\vec x}}{{d\xi }} =  - \hbar \sqrt {\frac{2}{m}} \vec \omega (\xi ) + \hbar \left( {\vec v\theta ( - \xi ) + \vec v'\theta (\xi )} \right).
\end{equation*}
When scattering in the high energy region, we can assume that the main contribution to the functional integral \eqref{3.8} is \textbf{the straight path} determined by initial and final momentum of the particle $\vec x(t) = t\left(\frac{\vec p}{m}\theta(-t) + \frac{\vec {p'}}{m}\theta(t)\right)$, that mean the contribute of the functional variables $\vec \omega(\eta)$  in the argument of potential in \eqref{3.8} is can be negligible.\\
Set $\vec \omega(\eta)\simeq 0$ in the \eqref{3.8} and if we called $\vec k,\vec {k'}$ respectively are the unit vectors in the direction of the momentum initial and the final of the particle, also set  $\alpha  = \left(\hbar^2 k\xi/m\right)= \hbar v\xi,$ we have $d\xi  =\frac{m}{\hbar^2 k} d\alpha = \frac{1}{\hbar v}d\alpha$. In this approximation for the scattering amplitude we get the expression below
\begin{equation}\label{3.9}
\begin{split}
f(\vec{k'},\vec k) =&- \frac{{4{\pi ^2}m}}{{{\hbar ^2}}}\int {\frac{{d\vec x}}{{{{\left( {2\pi } \right)}^3}}}V\left( {\vec x} \right){e^{i(\vec k - \vec{k'})\vec x}}}\times\\
&\int_0^1 d\lambda \exp \left\{ { - i\frac{\lambda }{{\hbar v}}\mathop \smallint \limits_{ - \infty }^\infty  V\left( {\vec x + \alpha \left[ {\vec k'\theta \left( \alpha  \right) + \vec k\theta \left( { - \alpha } \right)} \right]} \right)d\alpha } \right\}.
\end{split}
\end{equation}
We consider $\vec{k}$ towards the $Oz$ axis and small angle scattering. Then  $\vec k - \vec k' = \vec q \bot \vec k$ or  $\vec q \bot Oz$, $\vec q$ called ${q_\bot}$. Since there is no component of $z$ then $i\left( {\vec k - \vec k'} \right)\vec x = i{\vec q}{\vec b}$ (${\vec b}$ included $ x,y $). \\
Note that $\frac{1}{a}\left( {{e^a} - 1} \right) = \int\limits_0^1 {d\lambda {e^{\lambda a}}} $, where $a =  - \frac{i}{{\hbar v}}\mathop \smallint \limits_{ - \infty }^\infty  V\left( {\vec x + \alpha \left[ {\vec k'\theta \left( \alpha  \right) + \vec k\theta \left( { - \alpha } \right)} \right]} \right)d\alpha $.\\
And when $\vec{k}$ is oriented towards the $z$ - axis then the argument of $ V $ just changes the components with $z$ so $\int_{-\infty}^\infty V\left(\vec x + \alpha \left[ \vec{k'}\theta(\alpha)+ \vec k\theta(-\alpha)\right] \right)d\alpha  = \int_{- \infty}^\infty {V(x,y,z')dz'}$. Then we obtain
\begin{equation}\label{3.10}
\begin{split}
f\left( {\vec k,\vec k'} \right)=& - \frac{{4{\pi ^2}m}}{{{\hbar ^2}}}\frac{1}{{{{(2\pi)}^3}}}\int {dxdy\int {dzV(x,y,z){e^{i{{\vec q }{\vec b}}}}\frac{{{e^\alpha } - 1}}{{ - \frac{i}{{\hbar \nu }}\int\limits_{ - \infty }^{ + \infty } {dz'} V\left( {x,y,z'} \right)}}}} \\
 =&\frac{k}{{2\pi }}\int {{d^2}\vec x{e^{i{{\vec q}}{{\vec b}}}}\left( {{e^{ - \frac{i}{{\hbar \nu }}\int_{-\infty}^{+ \infty } {dz'} V(x,y,z')}} - 1} \right)}.
 \end{split}
\end{equation}
Equation \eqref{3.10} is the Glauber representation or also called the eikonal representation for scattering amplitude:
\begin{equation*}
f(\vec{k'},\vec k) = \frac{k}{{2\pi i}}\int {d^2}x{e^{i{{\vec q}}{{\vec b}}}}\left( {{e^{ - \frac{i}{{\hbar v}}\int_{- \infty}^\infty V(\vec x)dz}} - 1} \right).
\end{equation*}

\section{Scattering differential cross section for specific potential}
Using the eikonal representation for the scattering amplitude obtained in the previous section, we can find the differential scattering cross section for the scattering processes in specific external potential such as Yukawa potential and Gaussian potential.
\subsection{Yukawa potential}
The Yukawa potential has the form:
\begin{equation}\label{4.1}
V\left( r \right) = \frac{g}{r}{e^{ - \mu r}},
\end{equation}
where  $ g $ is a constant with dimension of energy,  $\mu $ also a constant.  We use equation \eqref{4.1} to calculate  the  scattering  phase [2]:
\[\chi (\vec b) =  - \frac{1}{{\hbar \nu }}\int\limits_{ - \infty }^\infty  {V\left( {\sqrt {{b^2} + {z^2}} } \right)dz}. \]
Substituting the Yukawa potential in eq.\eqref{4.1} into the above scattering phase, we have
\begin{equation}\label{4.2}
\chi (\vec b) = \frac{{1 + i\varepsilon }}{{\hbar \nu }}\int\limits_{ - \infty }^\infty  {V\left( {\sqrt {{b^2} + {z^2}} } \right)dz}  = \frac{{1 + i\varepsilon }}{{\hbar \nu }}\int\limits_{ - \infty }^\infty  {\frac{g}{r}{e^{ - \mu r}}dz}  = \frac{{2g}}{{\hbar \nu }}{K_0}\left( {\mu b} \right),
\end{equation}
where ${K_0}\left( {\mu b} \right) = \int_0^{+\infty} {\frac{{{e^{ - \mu \sqrt {{b^2} + {z^2}} }}}}{{\sqrt {{b^2} + {z^2}} }}dz} $ is the Mac Donal Fuction zeroth order (Modified Bessel function). Plugging the expression of the scattering phase \eqref{4.2} into the expression of the scattering amplitude, we obtain
\begin{equation}\label{4.3}
f\left( \theta  \right) = \frac{k}{i}\int\limits_0^\infty  {{J_0}} \left( {kb\theta } \right)\left\{ {{e^{i\chi \left( {\vec b} \right)}} - 1} \right\}bdb = \frac{k}{i}\int\limits_0^\infty  {{J_0}} \left( {kb\theta } \right)\left\{ {\exp \left[ {\frac{{2ig}}{{\hbar v}}{K_0}\left( {\mu b} \right)} \right] - 1} \right\}bdb.
\end{equation}
When $ b \to \infty  $  then $ {K_0}\left( {\mu b} \right) \to 0 $  the expression for the scattering amplitude becomes
\begin{equation}\label{4.4}
f\left( \theta  \right) = \frac{{2gk}}{{\hbar v}}(1 + i\varepsilon )\frac{1}{{{\mu ^2} - {k^2}{\theta ^2}}}.
\end{equation}
From the scattering amplitude \eqref{4.4} just found, we go to calculate the differential scattering cross-section. The result is
\begin{equation}\label{4.5}
\frac{{d\sigma }}{{d\Omega }} = {\left| {f\left( \theta  \right)} \right|^2} = {\left| {\frac{{2gk}}{{\hbar v}}(1 + i\varepsilon )\frac{1}{{{\mu ^2} - {k^2}{\theta ^2}}}} \right|^2} = 4{\left( {\frac{{gk}}{{\hbar v}}} \right)^2}{\left| {1 + i\varepsilon } \right|^2}\frac{1}{{{{\left( {{\mu ^2} - 4{k^2}{{\sin }^2}\frac{\theta }{2}} \right)}^2}}}.
\end{equation}
And the total scattering cross-section is also derived
\begin{equation}\label{4.6}
\sigma  = \frac{{16\pi {{(gk)}^2}{{(1 + i\varepsilon )}^2}}}{{{v^2}{\mu ^2}({\mu ^2} - 4{k^2})}}.
\end{equation}
\subsection{Gauss potential}
Gaussian potential has the form
\begin{equation}\label{4.7}
V\left( r \right) = g{e^{ - \alpha {r^2}}},
\end{equation}
where  $ g $  is the constant with energy dimensional,    $ \alpha $ is a real positive number. Similarly with Yukawa potential, we obtain the scattering phase
\begin{equation}\label{4.8}
\chi \left( {\overrightarrow b } \right) = \frac{{1 + i\varepsilon }}{{\hbar v}}\int\limits_{ - \infty }^\infty  {V\left( {\sqrt {{b^2} + {z^2}} } \right)} dz = \frac{g}{{\hbar v}}\int\limits_{ - \infty }^\infty  {{e^{ - \alpha {r^2}}}} dz = \frac{g}{{\hbar v}}(1 + i\varepsilon ){e^{ - \alpha {b^2}}}\sqrt {\frac{\pi }{\alpha }};
\end{equation}
From there we infer the scattering amplitude of the particle in the Gauss field as follow
\begin{equation}\label{4.9}
f\left( \theta  \right) = \frac{1}{{2\alpha }}(1 + i\varepsilon )\sqrt {\frac{\pi }{\alpha }} \frac{{gk}}{{\hbar v}}\exp \left( { - \frac{{{k^2}{\theta ^2}}}{{8\alpha }}} \right).
\end{equation}
The differential scattering cross-section and the total scattering cross-section are respectively
\begin{equation}\label{4.10}
\frac{{d\sigma }}{{d\Omega }} = \frac{\pi }{{4{\alpha ^3}}}{\left( {\frac{{gk}}{{\hbar v}}} \right)^2}\exp \left( { - \frac{{{k^2}{{\sin }^2}\frac{\theta }{2}}}{\alpha }} \right),
\end{equation}
\begin{equation}\label{4.11}
\sigma  = \frac{{{\pi ^2}}}{{2{\alpha ^2}}}{\left( {\frac{g}{{\hbar v}}} \right)^2}{(1 + i\varepsilon )^2}\left[ {1 - \exp \left( { - \frac{{{k^2}}}{\alpha }} \right)} \right].
\end{equation}
These expressions can be used to analyze current experimental data.
\section{Conclusion}
In this paper we study the problem of particle scattering in the external field in quantum mechanics by the functional integral method in straight line approximation, which is equivalent to the optical eikonal approximation. Glauber representation for the scattering amplitude of the external fast particles with small scattering angles was found through solving the Schrodinger equation by functional integral method. Scattering cross-sections of particles over specific external fields such as Yukawa and Gaussian potentials are the obtained. This approach will be used extensively to study scattering problems for quantum gravitational field theory.

\section{Acknowledgements}
We are grateful to thank Prof. Nguyen Suan Han for his suggestion of the problem and many useful comments. This work was supported funded by Viet- nam National Foundation for Science and Technology Development (NAFOSTED). DTH  is supported in part by the project 911 of Hanoi University of Science - VNU HN.
\newpage
\begin{center}
\textbf{APPENDIX}
\end{center}
\appendix
\section{The scattering amplitude in Born Approximation}\label{appA}
The Lippman Schwinger equation [8]
\begin{equation*}
f(\vec k,\vec k') =  - \frac{{4m{\pi ^2}}}{{{\hbar ^2}}}\left\langle {\vec k'\left| V \right|\vec k} \right\rangle  + \int {d\vec p} \left\langle {\vec k'\left| {V{G_0}} \right|\vec k} \right\rangle f(\vec p,\vec k).
\end{equation*}
The perturbation expansion solution of this equation will give us the scattering amplitude
\begin{equation}
f(\vec k',\vec k) =  - \frac{{4m{\pi ^2}}}{{{\hbar ^2}}}\left\langle {\vec k'\left| V \right|\vec k} \right\rangle  + \int {d\vec p} \left\langle {\vec k'\left| {V{G_0}} \right|\vec k} \right\rangle f(\vec p,\vec k)\tag{A.1}.
\end{equation}    				
Here
\begin{equation*}
\begin{split}
f^1(\vec k',\vec k) =& \frac{{4m{\pi ^2}}}{{{\hbar ^2}}}V(\vec {k'} - \vec k),\\
f^{(n + 1)}(\vec {k'},\vec k) =&  - \frac{{4m{\pi ^2}}}{{{\hbar ^2}}}\left\langle {\vec k'\left| {{t^{(n + 1)}}} \right|\vec k} \right\rangle  =  - \frac{{4m{\pi ^2}}}{{{\hbar ^2}}}\left\langle {\vec k'\left| {V{{({G_0}V)}^n}} \right|\vec k} \right\rangle {\rm{,  }}n \ge 1.
\end{split}
\end{equation*}
Series (A.1) gives us a simple interpretation by the graphs (see Figure 1). The line connecting the vertices corresponds to the propagation function ${G_0}$ (the factor $\frac{{2m}}{{{\hbar ^2}}}{({k^2} - {p^2} + i\varepsilon )^{ - 1}}$ appears in the momentum representation), and the wave line - is the Fourier image of potential $V(\vec p)$.\\
\begin{center}
    \begin{figure}[htp]
    \begin{center}
 \includegraphics[scale=.4]{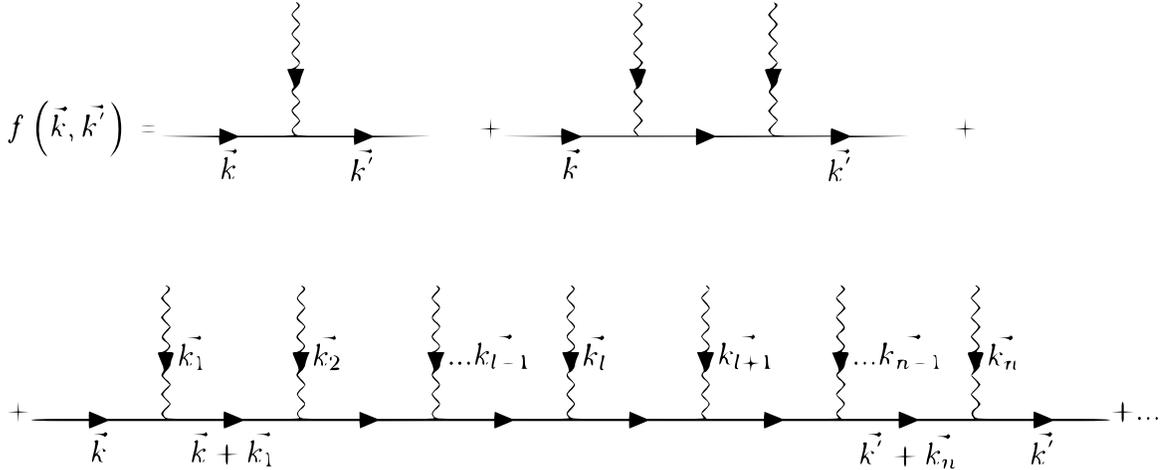}
    \end{center}
     \caption{Graphical representation of the Born series for potential scattering}
    \end{figure}
\end{center}
We consider the $(n+1)^{th}$ term of this series. Independent momentum variables can be impulses ${\vec k_i}(i = 1,2,...,n)$ as shown in Figure A.1 The contribution of this term to the scattering amplitude is equal to
\begin{equation}
{f^{(n + 1)}}(\vec k',\vec k) =  - \frac{{4m{\pi ^2}}}{{{\hbar ^2}}}{\left( {\frac{{2m}}{{{\hbar ^2}}}} \right)^n}\int {\prod_{i = 1}^n {\left( {d{{\vec k}_i}V({{\vec k}_i})} \right)} V\left( {\vec q - \sum_{i = 1}^n {{{\vec k}_i}} } \right)} {D_i}.\tag{A.2}
\end{equation}
where
\begin{equation*}
\begin{split}
D_l =&\left[k^2 - \left(\vec k + {\vec k}_1\right)^2 + i\varepsilon\right]^{- 1}\left[ {{k^2} - {{\left( {\vec k + {{\vec k}_1} + {{\vec k}_2}} \right)}^2} + i\varepsilon } \right]^{ - 1}...\\
&...{\left[ {{k^2} - {{\left( {\vec k + \sum_{i = 1}^l {{{\vec k}_i}} } \right)}^2} + i\varepsilon } \right]^{ - 1}}{\left[ {k{'^2} - {{\left( {\vec k' - \sum_{i = l + 1}^n {{{\vec k}_i}} } \right)}^2} + i\varepsilon } \right]^{ - 1}}...\\
&...{\left[ {k{'^2} - {{\left( {\vec k' - {{\vec k}_n}} \right)}^2} + i\varepsilon } \right]^{ - 1}},
\end{split}
\end{equation*}
$l$ is any integer that satisfies the condition $0 \le l \le n$.\\
Eq.(A.2) does not depend on which potential to be chosen in Figure 1. Therefore, we average by the position of potential separated in the middle of the rest after summing in eq.(A.2) follow $l$ from $0$ to $n$ and divide by $(n + 1)$
\begin{equation}
f^{(n + 1)}(\vec k',\vec k) =  - \frac{{4m{\pi ^2}}}{{{\hbar ^2}}}{\left( {\frac{{2m}}{{{\hbar ^2}}}} \right)^n}\int {\prod\limits_{i = 1}^n {\left( {d{{\vec k}_i}V({{\vec k}_i})} \right)} } V\left( {\vec q - \sum_{i = 1}^n {{{\vec k}_i}} } \right){\bar D_l}, \tag{A.3}
\end{equation}
with ${\bar D_l} = \frac{1}{{n + 1}}\sum_{l = 0}^n {{D_l}} $\\
We will use the eikonal approximation to compute the propagator function $G_0$ of scattered particles.We assume that the main contribution to the integral in eq.(A.2) is the integration taken on small momentum compared to initial and final momentums the particles. So to linearize the propagator function $G_0$ according to $\vec{k}_i$ we do the following substitution
\begin{equation}
{\left[ {{p^2} - {{\left( {\vec p + \sum_{i = 1}^m {{{\vec k}_i}} } \right)}^2} + i\varepsilon } \right]^{ - 1}} \to {\left[ { - 2\vec p\sum_{i = 1}^m {{{\vec k}_i}}  + i\varepsilon } \right]^{ - 1}}, \tag{A.4}
\end{equation}
where $\vec p = \vec k$ or $\vec k'$.
\begin{equation}
  {f^{(n + 1)}}(\vec k',\vec k) = - \frac{{4m{\pi ^2}}}{{{\hbar ^2}}}{\left( {\frac{{2m}}{{{\hbar ^2}}}} \right)^n}\frac{1}{{n + 1}}\sum\limits_{l = 0}^\infty  {\int {\prod_{i = 1}^l {\frac{{V({{\vec k}_i})d{{\vec k}_i}}}{{ - 2\vec k\sum\limits_{r = 1}^l {{{\vec k}_r} + i\varepsilon } }}} } }  \times. \tag{A.5}
\end{equation}
\begin{equation*}
 \times \prod_{j = l + 1}^n {\frac{{V({{\vec k}_j})d{{\vec k}_j}}}{{2\vec k'\sum\limits_{s = j}^n {{{\vec k}_s}}  + i\varepsilon }}} V\left( {\vec q - \sum_{i = 1}^n {{{\vec k}_i}} } \right)
\end{equation*}
Note that without averaging against $l$ in eq.(A.2), using the approximations in eq.(A.4) to eq.(A.2) we will receive the result that depend on the $l$ specific values. In other words, it depends on the arrangement of the momentum shown in figure 1.\\
Clearly, The expressions  $\prod\limits_{i = 1}^l {\frac{{V({{\vec k}_i})d{{\vec k}_i}}}{{ - 2\vec k\sum\limits_{r = 1}^l {{{\vec k}_r}}  + i\varepsilon }}} $;$\prod_{j = l + 1}^n {\frac{{V({{\vec k}_j})d{{\vec k}_j}}}{{2\vec k'\sum_{s = j}^n {{{\vec k}_s}}  + i\varepsilon }}} $ does not depend on the ordering possibilities for ${\vec k_i}(1 \le i \le l)$ and ${\vec k_j}(l + 1 \le j \le n)$ impulses. This arrangement may therefore be substituted for the following sum
\begin{equation}
\left\{\begin{array}{l}
\frac{1}{{l!}}\sum_{nepcet} {{{\vec k}_i}\frac{{V({{\vec k}_l})}}{{ - 2\vec k\sum_{r = 1}^l {{{\vec k}_r}}  + i\varepsilon }}} \\
\frac{1}{{(n - 1)!}}\sum_{nepcet} {{{\vec k}_j}\frac{{V({{\vec k}_j})}}{{2\vec k'\sum_{s = j}^n {{{\vec k}_s}}  + i\varepsilon }}}
\end{array} \right\} \tag{A.6}
\end{equation}
Substitute eq.(A.6) into eq.(A.5) and use identity in [20]
\begin{equation*}
  \sum_{nepcet} {\frac{1}{{C{'_1}}}\frac{1}{{C{'_1} + C{'_2}}}...\frac{1}{{C{'_1} + C{'_2} + ... + C{'_n}}}}  = \frac{1}{{{C_1}.{C_2}...{C_n}}},
\end{equation*}
here $\left\{ {C{'_1},C{'_2},...,C{'_n}} \right\}$ – is any arrangement of sequence $\left\{ {{C_1},{C_2},...,{C_n}} \right\}$, so we have
\begin{equation}
f^{(n + 1)}(\vec k',\vec k) =- \frac{{4m{\pi ^2}}}{{{\hbar ^2}}}{\left( {\frac{{2m}}{{{\hbar ^2}}}} \right)^n}\frac{1}{{n + 1}}\sum\limits_{l = 0}^\infty  {\frac{1}{{l!(n - l)!}}}.  \times \tag{A.7}
\end{equation}
\begin{equation*}
\times\int {\prod_{i = 1}^n {\frac{{V({{\vec k}_i})d{{\vec k}_i}}}{{ - 2\vec k{{\vec k}_i} + i\varepsilon }}.} \prod_{j = l + 1}^n {\frac{{V({{\vec k}_i})d{{\vec k}_i}}}{{2\vec k'{{\vec k}_i} + i\varepsilon }}.} } V\left( {\vec q - \sum_{i = 1}^n {{{\vec k}_i}} } \right)
\end{equation*}
Since $V\left( {\vec q - \sum_{i = 1}^n {{{\vec k}_i}} } \right) = \frac{1}{{{{(2\pi )}^3}}}\int {{e^{ - i\left( {\vec q - \sum\limits_{i = 1}^n {{{\vec k}_i}} } \right)\vec r}}} V(\vec r)d\vec r$ then eq.(A.7) can be expressed in form
\begin{equation}
f^{(n + 1)}(\vec k',\vec k)= - \frac{{4m{\pi ^2}}}{{{\hbar ^2}}}\int {\frac{{d\vec r}}{{{{(2\pi )}^3}}}{e^{ - i\vec q\vec r}}V(\vec r)\frac{1}{{n + 1}}}\times\tag{A.8}
\end{equation}
\begin{equation*}
\times \sum_{l = 0}^n {\frac{1}{{l!(n - l)!}}} {\left[ {U(\vec r,\vec k)} \right]^l}{\left[ {U({{\vec r}_j} - \vec k')} \right]^{n - l}}
\end{equation*}
where
\begin{equation}
U(\vec r,\vec k) = \frac{{2m}}{{{\hbar ^2}}}\int {\frac{{d\vec pV(\vec p){e^{i\vec r\vec p}}}}{{ - 2\vec k\vec p + i\varepsilon }}}. \tag{A.9}
\end{equation}
Summing up according to $l$ in eq.(A.8) can now be easily done
\begin{equation*}
{f^{(n + 1)}}(\vec k',\vec k) =  - \frac{{4m{\pi ^2}}}{{{\hbar ^2}}}\frac{1}{{(n + 1)!}}\int {\frac{{d\vec r}}{{{{(2\pi )}^3}}}{e^{ - i\vec q\vec r}}V(\vec r){{\left[ {U(\vec r,\vec k) + U({{\vec r}_j} - \vec k')} \right]}^n}}.
\end{equation*}
The total scattering amplitude is obtained with the following expression
\begin{equation}
f(\vec k',\vec k) = \sum_{n = 0}^\infty  {{f^{(n + 1)}}(\vec k',\vec k)}  =  - \frac{m}{{2\pi i{\hbar ^2}}}\int {d\vec r{e^{ - i\vec q\vec r}}V(\vec r)\frac{{\left[ {{e^{i\chi (\vec r,\vec k,\vec k')}} - 1} \right]}}{{^{\chi (\vec r,\vec k,\vec k')}}}},\tag{A.10}
\end{equation}
with
\begin{equation}
\chi (\vec r,\vec k,\vec k') =  - i\left[ {U(\vec r,\vec k) + U({{\vec r}_j} - \vec k')} \right].\tag{A.11}
\end{equation}
Using eq.(A.9) to transfer eikonal phase in eq.(A.11) into
\begin{equation}
\chi (\vec r) =  - \frac{1}{v}\int {d\xi V\left[ {\vec r + \xi \left( {\theta (\xi )\widehat {\vec k'} + \theta ( - \xi )\widehat {\vec k}} \right)} \right]},\tag{A.12}
\end{equation}
where $\widehat {\vec k'}$ and $\widehat {\vec k}$ are unit vectors directed towards the initial and final momentums of the particle, respectively.
Now, we can write eq.(A.10) in the form
\begin{equation}
f(\vec k',\vec k) =  - \frac{m}{{2\pi {\hbar ^2}}}\int {d\vec r} {e^{ - i\vec q\vec r}}V(\vec r)\int\limits_0^1 {d\lambda}
 \times \exp \left\{ { - \frac{{i\lambda }}{{\hbar v}}\int\limits_{ - \infty }^{ + \infty } {d\xi V} \left[ {\vec r + \xi \left( {\theta (\xi )\widehat {\vec k'} + \theta ( - \xi )\widehat {\vec k}} \right)} \right]} \right\}.\tag{A.13}
\end{equation}
The only difference (A.13) with Shiv's formula for the large angle scattering amplitude is that the integral according to $d\lambda $, it absent in eq.(4.1). In the case of small angle scattering, it is easy to change. formula (A.13) to the eikonal form (3.8). To do that just put  eikonal phase in eq.(A.12) $\widehat {\vec k'} = \widehat {\vec k}$ and $\vec q \bot \vec k$. The z-axis is normally oriented in the $\vec k$. After integrating according to dz in eq.(A.13) we obtained
\begin{equation*}
f(\vec k',\vec k) = \frac{k}{{2\pi i}}\int {{d^2}{b}} {e^{ - i{{\vec q}}{{\vec b}}}}\left\{ {\exp \left( { - \frac{i}{{\hbar \nu }}\int_{- \infty }^{ + \infty } {V(\vec r)} dz} \right) - 1} \right\}.
\end{equation*}
\section{Green function of the complete Schrodinger equation and scattering amplitude [19]}
Together with the Green function of the free Schrodinger equation we can consider the complete Schrodinger's equation.
\begin{equation}
(E - {H_0} + i\varepsilon )G(\vec r,\vec r') = (E - {H_0} - V + i\varepsilon )G(\vec r - \vec r').\tag{B.1}
\end{equation}
The total Green function will contain all information about the quantum system. Thanks to this function we can find the energy spectrum of the system, the wave function, the scattering amplitude. This relation can be established using operator notation.
Thanks to eqs.(2.5) and (B.3) $G - {G_0} $ can be expressed in the following form
\begin{equation}
 \left\{ \begin{array}{l}
G - {G_0} = {(E - {H_0} - V + i\varepsilon )^{ - 1}} - {(E - {H_0} + i\varepsilon )^{ - 1}} = {G_0}VG\\
{\rm{           }} = {G_0}V(G - {G_0} + {G_0}) = {G_0}V{G_0} + {G_0}V{G_0}{G_0}^{ - 1}(G - {G_0})
\end{array} \right\} \tag{B.2} 		
\end{equation}
Left and right multiply equation (B.4) by $ {G_0}^{ - 1} $, we get
\begin{equation*}
{G_0}^{ - 1}(G - {G_0}){G_0}^{ - 1} = V + V{G_0}{G_0}^{ - 1}(G - {G_0}){G^{ - 1}} .
\end{equation*}
From here, the quantity \[{G_0}^{ - 1}(G - {G_0}){G_0}^{ - 1}\] only satisfies the equation, as well as the scattering operator t (eq.(B.2)), thus
$$t =  - {G_0}^{ - 1}(G - {G_0}){G_0}^{ - 1}.$$.
The scattering amplitude in this case is determined by
\begin{equation}
\left\{\begin{array}{c}
f(\vec k,\vec k') =  - \frac{{4{\pi ^2}m}}{{{\hbar ^2}}}\left\langle {\vec k'\left| {{G_0}^{ - 1}(G - {G_0}){G_0}^{ - 1}\left| {\vec k} \right.} \right.} \right\rangle \\
 =  - \frac{{4{\pi ^2}m}}{{{\hbar ^2}}}\left( {E - \frac{{{\hbar ^2}\vec k{'^2}}}{{2m}} + i\varepsilon } \right)\left\langle {\vec k'\left| {G - {G_0}\left| {\vec k} \right.} \right.} \right\rangle \left( {E - \frac{{{\hbar ^2}{{\vec k}^2}}}{{2m}} + i\varepsilon } \right)
\end{array}\right\},\tag{B.3}
 \end{equation}
here $E = \frac{{{\hbar ^2}{{\vec k}^2}}}{{2m}} = \frac{{{\hbar ^2}\vec k{'^2}}}{{2m}}$.\\
This is possible, if we use the Green function of the free Schrodinger equation $G_0(\vec{r},\vec{r'})$, which satisfies the equation below.
\begin{equation}
\left( {E - {H_0} + i\varepsilon } \right){G_0}(\vec r,\vec r') = \left( {E + \frac{{{\hbar ^2}}}{{2m}}\Delta  + i\varepsilon } \right){G_0}(\vec r,\vec r') = {\delta ^{(3)}}(\vec r,\vec r').\tag{B.4}
\end{equation}
The ${G_0}(\vec r,\vec r')$ function has the following form [6]
\begin{equation}
\left\{ \begin{array}{c}
{G_0}(\vec r,\vec r') = {\left( {E - {H_0} + i\varepsilon } \right)^{ - 1}}{\delta ^{(3)}}(\vec r,\vec r') = \frac{1}{{{{(2\pi )}^3}}}\int {\frac{{{e^{i\vec q(\vec r - \vec r')}}}}{{E - \frac{{{\hbar ^2}{q^2}}}{{2m}} + i\varepsilon }}} d\vec q\\
 = \frac{{2m}}{{{\hbar ^2}}}\int {\frac{{d\vec q}}{{{{(2\pi )}^3}}}} \frac{{{e^{i\vec q(\vec r - \vec r')}}}}{{{k^2} - {q^2} + i\varepsilon }} =  - \frac{1}{{4\pi }}\frac{{2m}}{{{\hbar ^2}}}\frac{{{e^{i\vec q(\vec r - \vec r')}}}}{{\left| {\vec r - \vec r'} \right|}}
\end{array}\right.\tag{B.5}
\end{equation}

\end{document}